\documentclass[fp,twocolumn]{jpsj3}
\usepackage{txfonts}
\usepackage{color}

\title{
Haldane Gaps of Large-$S$ Heisenberg Antiferromagnetic Chains \\ 
and Asymptotic Behavior
}

\author{Hiroki Nakano$^1$\thanks{hnakano@sci.u-hyogo.ac.jp}, 
Norikazu Todoroki$^{2}$, 
 and T\^oru Sakai$^{1,3}$}
\inst{$^1$
Graduate School of Material Science, 
University of Hyogo,
Kamigori, 
Hyogo 678-1297, Japan \\
$^2$
Chiba Institute of Technology, Narashino, Chiba 275-0023, Japan \\
$^3$
National Institutes for Quantum and Radiological Science and Technology, 
SPring-8, 
Sayo, Hyogo 679-5148, Japan %\\
}

\abst{
The one-dimensional Heisenberg antiferromagnets 
of large-integer-$S$ spins 
are studied; their Haldane gaps are estimated 
by the numerical diagonalization method for $S=5$ and 6. 
We successfully obtain 
a monotonically increasing sequence 
of finite-size energy difference data
corresponding to the Haldane gaps 
from the huge-scale parallel calculations of diagonalization 
under the twisted boundary condition 
and create
a monotonically decreasing sequence 
within the range of system sizes treated in this study 
from the monotonically increasing sequence. 
Consequently,  
the gaps for $S=5$ and 6 are estimated to be 
$0.000050 \pm 0.000005$ 
and 
$0.0000030 \pm 0.0000005$, respectively. 
The asymptotic formula of the Haldane gap for $S\rightarrow\infty$ 
is examined from the new estimates to determine 
the coefficient in the formula more precisely. 
}

\begin{document}
\maketitle

\section{Introduction}

The Haldane gap 
\---
the energy gap between the unique ground state and the first excited state 
for the integer-$S$ Heisenberg antiferromagnets in one dimension 
\---
is now well known; however, 
the presence of the gap surprised many condensed-matter physicists 
when it was originally conjectured in Refs.~\ref{Haldane1} and \ref{Haldane2} 
by mapping the Heisenberg chain to the nonlinear $\sigma$ model. 
Various investigations concerning whether the gap is present or absent 
have been carried out 
since Refs.~\ref{Haldane1} and \ref{Haldane2} were published. 
Currently, the existence of a nonzero gap is widely believed. 
Extensive studies concerning this phenomenon 
have contributed considerably to our understanding 
of the various properties of quantum spin systems. 

To date, various approaches have been attempted 
to estimate the magnitude of the gap. 
In particular, the Haldane gap for $S=1$ 
was studied\cite{Nightingale1986,TSakai1900} 
soon after the conjecture 
because the $S=1$ gap is larger than those for larger $S$. 
It is now known that 
the density matrix renormalization group (DMRG)\cite{White_DMRG_PRB1993}, 
quantum Monte Carlo (QMC)\cite{TodoKato_PRL2001}, and 
numerical diagonalization (ND)\cite{HNakano_HaldaneGap_JPSJ2009} calculations 
give estimates that agree with each other to
five decimal places: 
$\Delta/J \sim 0.41048$, 
where $\Delta$ and $J$ represent 
the gap value and the strength of the interaction defined later, 
respectively. 
For $S=2$, 
after various studies\cite{Wang_PRB1999,TodoKato_PRL2001,
HUeda_PRB2011,HN_TSakai_S2HaldaneGap_JPSJ2018}, 
the gap estimates from the three approaches 
agree with each other within errors; 
$\Delta/J \sim 0.089$.  

For cases of even larger $S$, however, 
it is more difficult to estimate the gap values. 
Therefore, the number of studies for $S$ larger than $2$ 
is much smaller. 
The first report of the $S=3$ case is Ref.~\ref{TodoKato_PRL2001},   
which reported $\Delta/J = 0.01002 \pm 0.00003$ 
from a QMC calculation. 
This estimate was confirmed 
by an ND study\cite{HNakano_HaldaneGap_JPSJ2009} reporting 
$\Delta/J = 0.0092 \pm 0.0010$.  
The ND study also provided the gap estimates for even larger $S$:  
$\Delta/J = 0.00072 \pm 0.00009$ for $S=4$ 
and    
$\Delta/J = 0.000047 \pm 0.000010$ for $S=5$. 
This ND study was the first report for $S=4$ and $S=5$.  
Among the estimates, 
a very recent QMC calculation\cite{Todo_Matsuo_Shitara_CPC2019} gave 
$\Delta/J = 0.000799 \pm 0.000005$ for $S=4$, 
which agrees with the estimate by the ND calculation. 
Unfortunately, no other approaches have successfully 
estimated the gap value for $S$ larger than $4$ 
to the best of our knowledge. 
The estimation of the Haldane gap for large $S$ 
is still one of the most challenging issues 
in condensed-matter physics, particularly, 
from the viewpoint of computational statistical physics. 

One reason why it is quite difficult to estimate the gap values numerically 
is that a high-cost calculation is necessary 
to treat considerably large systems. 
The requirement is strongly related to the fact that 
the gap values for large $S$ are extremely small although they are nonzero. 
It is notable that 
the ND study\cite{HNakano_HaldaneGap_JPSJ2009} succeeded 
in the estimation 
by using the twisted boundary condition 
even though the ND method can only treat clusters that are 
much smaller than those treated by the QMC and DMRG calculations. 
Therefore, the validity of the method used 
in Ref.~\ref{HNakano_HaldaneGap_JPSJ2009} should be examined 
from various viewpoints. 

Concerning the Haldane gaps for large $S$, 
the original studies by Haldane\cite{Haldane1,Haldane2} 
derived their asymptotic formula as follows: 
\begin{equation}
\Delta(S)/J= \beta |{\bf S}|^2 
\exp (-\pi |{\bf S}|), 
\label{asymptotic_formula}
\end{equation}
for $S\rightarrow\infty$,
where $|{\bf S}|$ represents the amplitude of each spin
included in the target Heisenberg chain. 
However, the coefficient $\beta$ cannot be determined 
only by Haldane's argument. 
To determine $\beta$, 
numerical approaches are required for sufficiently large $S$. 
The ND calculations in Ref.~\ref{HNakano_HaldaneGap_JPSJ2009} 
gave an estimate of 
\begin{equation}
\beta = 12.8 \pm 1.5 .
\label{estimate_beta_2009}
\end{equation}
No other estimates are known to the best of our knowledge. 
If one combines Eq.~(\ref{asymptotic_formula}) and the estimate of $\beta$, 
one can infer the gap value for even larger $S$.
For $S=6$, for example, one can easily predict 
\begin{equation}
\Delta/J = 0.0000030 \pm 0.0000004 ,
\label{S6prediction}
\end{equation}
from Eq.~(\ref{asymptotic_formula}). 
If one can directly estimate the gap value for $S=6$, 
it will be possible to compare the estimate 
with Eq.~(\ref{S6prediction}). 
If the estimate agrees with Eq.~(\ref{S6prediction}),  
it will suggest 
the validity of the method used in Ref.~\ref{HNakano_HaldaneGap_JPSJ2009}. 

\begin{table}[tb]
\caption{Finite-size gaps for the $S=5$ case 
under the twisted boundary condition. 
The dimensions of the calculated subspace of $M=0$ and 
the ground-state energies per site for each $N$ 
are also presented.
}
\label{gap_S5_table}
\begin{tabular}{r|r|c|c}
\hline
$N$ & dimension & $-E_0/N$ & $10^{5} \Delta_{N}/J$ 
\\
\hline
 4 & 891 & 26.48591940774 & 0.59777  \\  
 6 & 88913 & 26.69461445758 & 2.06995  \\  
 8 & 9377467 & 26.76390333054 & 3.12913  \\  
10 & 1018872811 & 26.79535044080 & 3.79423  \\  
12 & 112835748609 & 26.81227098785 & 4.22111  \\  
14 & 12663809507129 & 26.82241910365 & 4.50720  \\  
\hline
\end{tabular}
\end{table}

Under these circumstances, 
the purpose of this study is to verify 
the method used in Ref.~\ref{HNakano_HaldaneGap_JPSJ2009} 
from the following two aspects. 
The first one is to obtain a direct estimate of the gap value for $S=6$ 
by the method used in Ref.~\ref{HNakano_HaldaneGap_JPSJ2009}. 
The additional estimate allows it to be compared 
with the prediction (\ref{S6prediction}). 
The second aspect is to obtain finite-size energy gaps 
for $S=5$ with clusters that were not treated 
in Ref.~\ref{HNakano_HaldaneGap_JPSJ2009}. 
Although in Ref.~\ref{HNakano_HaldaneGap_JPSJ2009}, 
the authors were able to treat 
only clusters up to 10 sites,
in this study, we additionally report 
results for clusters with 12 and 14 sites;  
we will confirm that the additional results show 
the common behavior of the results up to 10 sites 
and give an estimate that is more precise 
than that in Ref.~\ref{HNakano_HaldaneGap_JPSJ2009}. 

This paper is organized as follows. 
In the next section, the model Hamiltonian will be introduced. 
Our numerical method will also be explained. 
The third section is devoted 
to the presentation and discussion of our results. 
We will first treat the case of $S=5$. 
Next, we will study the case of $S=6$.  
From the additional gap values for $S=5$ and $S=6$, 
finally, we will estimate the coefficient $\beta$ more precisely. 
In the final section, 
we will summarize our results and give some remarks. 

\section{Model Hamiltonian and Numerical Method}

The Hamiltonian studied here is given by 
%\begin{eqnarray}
\begin{equation}
{\cal H}
%&=& 
=
\sum_{i} J
\mbox{\boldmath $S$}_{i}\cdot\mbox{\boldmath $S$}_{i+1} 
, 
\label{Hamiltonian}
%\end{eqnarray}
\end{equation}
where $\mbox{\boldmath $S$}_{i}$ 
represents the spin-$S$ spin operator at site $i$. 
In this study, we particularly focus our attention 
on the cases of $S=5$ and $S=6$.
We consider the case of an isotropic interaction 
in spin space in this study. 
The label of a spin site is represented by $i$, 
which should be an integer. 
The number of spin sites is denoted by $N$, 
which is assumed to be an even integer. 
Energies are measured in units of $J$; 
hereafter, we set $J=1$, which indicates that 
the system is an antiferromagnet. 
We treat finite-size clusters with system size $N$ 
under the twisted boundary condition.
This condition is given by
\begin{equation}
  S^{x}_{N+1}=-S^{x}_{1}, \ 
  S^{y}_{N+1}=-S^{y}_{1}, \ 
  S^{z}_{N+1}= S^{z}_{1}, 
\label{TBC}
\end{equation}
which should be noted by the difference from 
the periodic boundary condition given by
$\mbox{\boldmath $S$}_{N+1}=\mbox{\boldmath $S$}_{1}$. 
Owing to the twisted boundary condition, 
the system size $N$ should satisfy $N\ge 4$.
Note also that the twisted boundary condition 
in studies of quantum spin systems is not so strange 
because the condition is effectively used in studies 
using level-spectroscopy analysis\cite{Kitazawa_TBC_LSS,
Kitazawa_Nomura1997,Nomura_Kitazawa1998,Tonegawa_etal_2011}.  
In particular, in Ref.~\ref{Kitazawa_Nomura1997}, 
the authors used the twisted boundary condition 
to study the $S=1$ one-dimensional antiferromagnet 
with bond alternation 
and 
successfully determined the gapless point between 
the two gapped phases, namely, the Haldane phase 
and the dimer phase. 
To find the gapless point in finite-size systems, 
in this reference, the authors searched for a level-crossing point 
of the ground state of the system 
under the twisted boundary condition. 
This means that a gapless case is realized in finite-size systems 
as a situation of the doubly degenerate ground state. 
The degeneracy indicates that the energy difference 
between the two states vanishes. 
One thus finds that the twisted boundary condition 
can appropriately capture such a gapless case. 
In Ref.~\ref{HNakano_HaldaneGap_JPSJ2009}, on the other hand, 
the authors applied the twisted boundary condition in gapped cases 
to show that the condition can contribute to the gap estimation. 
The standing position of this paper 
is to confirm the validity of the method used 
in Ref.~\ref{HNakano_HaldaneGap_JPSJ2009}. 
The reason why the twisted boundary condition is used here 
will also be mentioned 
when our practical results are presented 
in the next section. 

\begin{figure}[tb]
\begin{center}
\includegraphics[width=8cm]{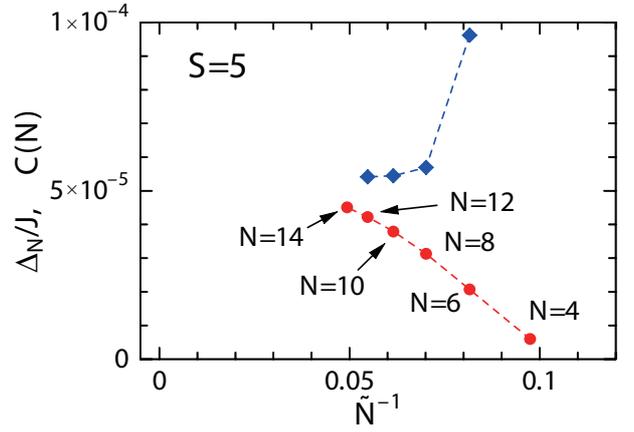}
\end{center}
\caption{(Color) 
Finite-size gaps $\Delta_N$ for $S=5$ under the twisted boundary condition 
are indicated by red closed circles. 
We determine a normalized system size $\tilde{N}$ defined as $N+N_{0}$ 
so that the three data for $N=4$, 6, and 8 reveal a linear dependence: 
$N_{0}=6.25840221$. 
Blue closed diamonds represent
$C(N)$. 
Equation~(\ref{fit_upper_bound_no_acceleration}) gives $C(N)$ 
from the finite-size gaps of system sizes $N-2$, $N$, and $N+2$. 
}
\label{fig1}
\end{figure}

We carry out our numerical diagonalizations 
based on the Lanczos algorithm 
to obtain the lowest energies of ${\cal H}$ 
in the subspace characterized by $\sum _j S_j^z=M$. 
Note here that the $z$-axis 
is taken as the quantized axis of each spin.  
ND calculations can treat quantum effects 
without any approximations and 
provide us with very precise results within numerical errors. 
Thus, one can obtain reliable information about the system. 
We focus our attention on the lowest energy excitation;
therefore, our calculations are carried out for $M=0$. 
The lowest energy and the first excited energy for a given $N$  
are denoted by $E_{0}$ and $E_{1}$, respectively. 
We evaluate the energy difference given by
\begin{equation}
\Delta_{N}=E_{1} - E_{0} ,  
\label{gap_determining_finite_N}
\end{equation}
for each $N$. 

Some of the Lanczos diagonalizations were carried out 
using an MPI-parallelized code that was originally 
developed in Ref.~\ref{HNakano_HaldaneGap_JPSJ2009}. 
The usefulness of our program was confirmed in large-scale 
parallelized calculations\cite{HNakano_kgm_gap_JPSJ2011,
HNakano_s1tri_LRO_JPSJ2013,HN_TSakai_kgm_1_3_JPSJ2014,
HN_TSakai_kgm_S_JPSJ2015,HN_YHasegawa_TSakai_dist_s
huriken_JPSJ2015,HN_TSakai_dist_tri_JPSJ2017,
HN_TSakai_tri_NN_JPSJ2017,HN_TSakai_kgm45_JPSJ2018,
YHasegawa_HN_TSakai_dist_shuriken_PRB2018,
TSakai_HN_ICM018,HN_TSakai_S2HaldaneGap_JPSJ2018,
Jeschke2019}. 
Note here that the largest scale
calculations in this study 
have been carried out using either the K computer or Oakforest-PACS. 
We carry out our calculations up to $N=14$ for $S=5$
and up to $N=12$ for $S=6$. 
Among our calculations, 
the case of $N=14$ for $S=5$ is the largest\cite{comment_large_scale_job} 
with respect to the dimension of the calculated subspace,
which is 12,663,809,507,129.
Note here that this dimension is larger than
5,966,636,799,745 for the case of $N=20$ for $S=2$
in which Lanczos diagonalization was successfully carried out 
in Ref.~\ref{HN_TSakai_S2HaldaneGap_JPSJ2018}. 

\begin{table}[tb]
\caption{Finite-size gaps for the $S=6$ case 
under the twisted boundary condition. 
The dimensions of the calculated subspace of $M=0$ and 
the ground-state energies per site for each $N$ 
are also presented.
}
\label{gap_S6_table}
\begin{tabular}{r|r|c|c}
\hline
$N$ & dimension & $-E_0/N$ & $10^{6} \Delta_{N}/J$ 
\\
\hline
 4 &         1469 & 37.77880919592 & 0.25442  \\  
 6 &       204763 & 38.02791744421 & 1.11540  \\  
 8 &     30162301 & 38.11057649245 & 1.81274  \\  
10 &   4577127763 & 38.14808361999 & 2.26870  \\  
12 & 707972099627 & 38.16826289951 & 2.56609   \\  
\hline
\end{tabular}
\end{table}

\section{Results and Discussion}

\subsection{Case for $S=5$}

Now, we study our results for $S=5$; 
our numerical results under the twisted boundary condition 
are presented in Table~\ref{gap_S5_table}. 
One can observe that 
the finite-size gap $\Delta_N$ is
monotonically increasing 
with respect to $N$. 
This monotonic increase 
is a decisive merit 
of using the twisted boundary condition. 
As mentioned in Ref.~\ref{HNakano_HaldaneGap_JPSJ2009}, 
if we use the periodic boundary condition, 
each finite-size gap has a significantly large magnitude and 
is monotonically decreasing with increasing $N$. 
From such a data sequence, it is extremely difficult 
to extrapolate the sequence to the thermodynamic limit 
and to find whether the nonzero gap is present or absent. 
On the other hand, 
a data sequence with a monotonic increase 
can easily enable us to find that the gap opens. 
In Ref.~\ref{HNakano_HaldaneGap_JPSJ2009}, 
the dependence for $S=5$ was observed up to 10 sites. 
In this study, we successfully clarified that 
our additional results for $N=12$ and 14 maintain 
the monotonic increase. 

Next, we analyze our results and 
produce another sequence that is
monotonically decreasing 
within the range of $N$ treated here 
and that approaches 
the gap in the thermodynamic limit 
$N\rightarrow\infty$ according to the method 
used in Ref.~\ref{HNakano_HaldaneGap_JPSJ2009},  
which employs a two-step procedure. 

The first step is to draw a plot of $\Delta_N$ 
as a function of $1/\tilde{N}$ instead of raw $N$, 
where we introduce a renormalized system size $\tilde{N}$ 
defined as $N+N_{0}$ so that 
the three initial data for $N=4$, 6, and 8 reveal a linear dependence 
in the plot of $\Delta_N$. 
The result for $S=5$ is depicted in Fig.~\ref{fig1}. 
If we skip the first step and 
draw a usual plot of $\Delta_N$ versus $1/N$, 
the $1/N$ dependence of $\Delta_N$ shows
concave upward behavior 
for small sizes. 
In Ref.~\ref{HNakano_HaldaneGap_JPSJ2009}, 
the authors reported that 
the data up to $N=10$, 
which was the maximum of $N$ in this reference, 
do not show the
convex upward 
behavior in the plot of $\Delta_N$ versus $1/N$. 
Under such a situation, it was difficult 
to obtain an appropriate extrapolated value for $\Delta_N$ 
from the analysis using the plot of $\Delta_N$ versus $1/N$, 
which was why $\Delta_N$ versus $1/\tilde{N}$
was plotted in Ref.~\ref{HNakano_HaldaneGap_JPSJ2009}. 
In this paper, on the other hand, we additionally report 
data for $N=12$ and 14,
in which convex upward behavior is observed
for large system sizes in the plot of $\Delta_{N}$ versus $1/N$. 
Since the purpose of this paper is to examine the validity 
of the method used in Ref.~\ref{HNakano_HaldaneGap_JPSJ2009}, 
we followed the same procedure 
and will later discuss the cases of using $\tilde{N}$ or raw $N$. 

\begin{figure}[tb]
\begin{center}
\includegraphics[width=8cm]{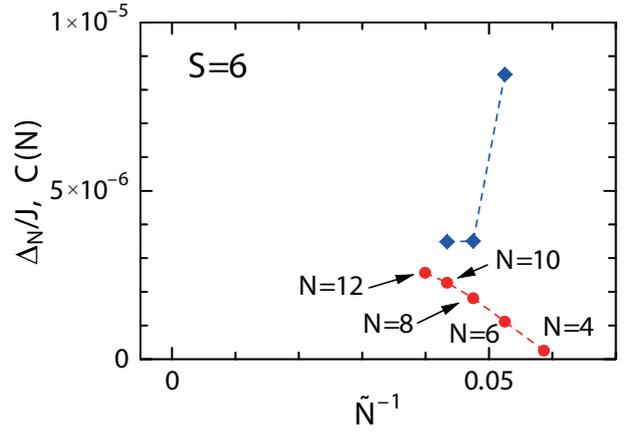}
\end{center}
\caption{(Color) 
Finite-size gaps $\Delta_N$ for $S=6$ under the twisted boundary condition 
are indicated by red closed circles. 
We determine a normalized system size $\tilde{N}$ defined as $N+N_{0}$ 
so that the three data for $N=4$, 6, and 8 reveal a linear dependence: 
$N_{0}=13.0448842$. 
Blue closed diamonds represent
$C(N)$. 
Equation~(\ref{fit_upper_bound_no_acceleration}) gives $C(N)$ 
from the finite-size gaps of system sizes $N-2$, $N$, and $N+2$. 
}
\label{fig2}
\end{figure}

The second step is to create a decreasing sequence 
from $\Delta_N$, which is an increasing sequence in the plot 
of $\Delta_N$ as a function of $\tilde{N}$. 
We focus our attention on three neighboring data points of system sizes 
$N$$-$2, $N$, and $N$$+$2 in Fig.~\ref{fig1}. 
When we apply the fitting curve of 
\begin{equation}
y = C + D x^{E},
\label{fit_upper_bound_no_acceleration}
\end{equation}
to the neighboring three data points, 
we can determine the parameters $C$, $D$, and $E$ 
uniquely for a given $N$. 
Thus, we use $C(N)$, $D(N)$, and $E(N)$ hereafter. 
Note here that $E(N=6)$ is necessarily unity 
owing to the above first step. 
The result for $C(N)$ is also depicted 
at the corresponding $\tilde{N}$ in Fig.~\ref{fig1}.  
One can observe in Fig.~\ref{fig1} that 
$C(N)$ is
monotonically decreasing when $N$ is increased 
within the range of $N$ treated here.
The sequence $C(N)$ is approaching the gap value 
in the thermodynamic limit from the side that is opposite to 
that of $\Delta_{N}$. 
Note that $C(N)$ approaches the gap value
in the thermodynamic limit 
regardless of whether one uses $\tilde{N}$ or $N$. 
The behavior of approaching the gap value from the opposite side 
suggests that
$C(N)$ for the largest system size 
is the closest to the gap value that we want to know finally; 
$C(N)$ for the largest system size in the present paper is 
$\sim$0.0000542. 
Therefore, 
we obtain 
\begin{equation}
\Delta(S=5)/J = 0.000050 \pm 0.000005, 
\label{gap_s5_new}
\end{equation}
as a new estimate whose error is smaller 
than that in Ref.~\ref{HNakano_HaldaneGap_JPSJ2009}. 
Consequently, we find that 
the method used in Ref.~\ref{HNakano_HaldaneGap_JPSJ2009} 
can be applied for the additional data for $N=12$ and $N=14$, 
which are larger 
than those treated in Ref.~\ref{HNakano_HaldaneGap_JPSJ2009}.  
Note here that 
$C(N)$ depends on whether one uses 
the plot of $\Delta_{N}$ versus $1/\tilde{N}$ or 
that versus $1/N$. 
If one uses the plot of $\Delta_{N}$ versus $1/N$, 
one obtains $C(N)$ for the largest system size in this paper 
to be $\sim$0.0000595. 
The comparison of the two results for $C(N)$ from $\tilde{N}$ or $N$ 
suggests that the usage of $\tilde{N}$ 
causes the error of an estimate for the gap value to become smaller. 

\subsection{Case for $S=6$}

Next, we study the case of $S=6$; 
our numerical results under the twisted boundary condition 
are presented in Table~\ref{gap_S6_table}. 
This study is the first report giving
finite-size gaps for $S=6$ to the best of our knowledge. 
One can also observe that 
the finite-size gap $\Delta_N$ is
monotonically increasing 
with respect to $N$ 
and find that the behavior of
the monotonic increase 
is certainly maintained for $S=6$. 

Let us analyze
the monotonically increasing $\Delta_N$ 
and obtain 
a monotonically decreasing sequence
that is approaching 
the gap value in the thermodynamic limit 
according to the same method as in the $S=5$ case;  
the results are depicted in Fig.~\ref{fig2}. 
We successfully obtain a monotonically decreasing $C(N)$ 
within the range of $N$ treated here. 
We finally obtain an estimate for 
the gap value for $S=6$ to be 
\begin{equation}
\Delta(S=6)/J =
0.0000030 \pm 0.0000005. 
\label{gap_s6_new}
\end{equation}
One finds that 
the present Eq.~(\ref{gap_s6_new}) agrees with 
the prediction (\ref{S6prediction}). 
The agreement strongly indicates the validity 
of the method used in Ref.~\ref{HNakano_HaldaneGap_JPSJ2009} 
and this study.  
Our estimate for $S=6$ will be 
investigated in the future if other methods become available.

\subsection{Asymptotic behavior}

Now, we examine 
the asymptotic formula of Eq.~(\ref{asymptotic_formula})
for the Haldane gap
for $S\rightarrow\infty$ from our estimate of the $S=5$ and 6 gaps. 
To do this, 
we introduce new parameters $x = S^{-1}$ and 
$y = S^{-1} \log ( S^2 J/\Delta(S))$ 
when the amplitude of each spin is taken to be $|{\bf S}|=S$ 
in this analysis. 
The asymptotic formula (\ref{asymptotic_formula}) 
is rewritten as 
\begin{equation}
y = \pi - x \log \beta. 
\label{analysis4asymptotic}
\end{equation}

\begin{figure}[tb]
\begin{center}
\includegraphics[width=8cm]{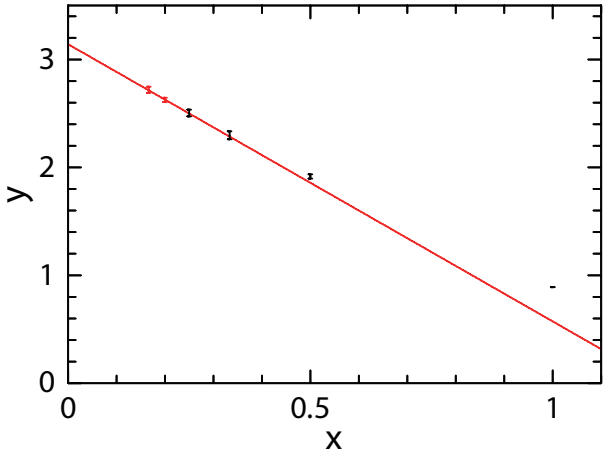}
\end{center}
\caption{(Color) 
Analysis of our estimates of Haldane gaps 
for $S=5$ and 6 denoted by red error bars. 
The Haldane gaps up to $S=4$, 
which are shown by black error bars, 
are taken from Ref.~\ref{HNakano_HaldaneGap_JPSJ2009}.  
The red dotted line corresponds to $\beta\sim 13.0$ 
obtained from the linear fitting of $S=5$ and 6. 
}
\label{asympt4j}
\end{figure}

Let us input the obtained estimates of the Haldane gaps 
to $\Delta(S)$ in $y$ and plot the $x$ dependence of $y$. 
The result is depicted in Fig.~\ref{asympt4j}. 
One can find linear behavior for finite but large $S$ 
up to $S=6$. 
We have fitted our data for $S=5$ and 6 
using the straight line (\ref{analysis4asymptotic}); 
the best fit is produced by 
$\beta = 13.0 \pm 1.2$. 
The good linear behavior suggests that 
the asymptotic formula (\ref{asymptotic_formula}) holds well 
for large $S$. 

\section{Summary and Remarks}

We have studied 
the Haldane gaps of the integer-$S$ Heisenberg antiferromagnetic chain model 
when $S$ is large 
by the Lanczos diagonalization method. 
We have successfully obtained an estimate for $S=5$ 
that is more precise than that in the previous study. 
We have also succeeded in obtaining an estimate for $S=6$ 
for the first time. 
This study allows us to confirm 
the validity of numerical diagonalization calculations 
under the twisted boundary condition and 
the analysis giving
an estimate for the gap value in the thermodynamic limit. 
Cases for even larger $S$ will be studied in future works. 
Such studies are expected to contribute much 
to our fundamental understanding of quantum magnetism 
and further development of our numerical techniques. 

Finally, let us give some comments concerning experiments. 
To date, experimental studies of the Haldane gaps 
have also been carried out. 
For $S=1$, 
NEMP\cite{Renard_NEMP1987,Katsumata_NEMP1989,Ajiro_NEMP1989} 
is a famous achievement as a good candidate material. 
In recent years, some materials have been reported 
for $S=2$\cite{SShinozaki_PRB2018,YIwasaki_PRB2018}. 
Unfortunately, there are no experimental reports for $S=3$ 
to the best of our knowledge. 
Since the gap magnitudes for $S$ larger than $2$ 
are quite small, 
it is necessary to remove 
effects from interactions other 
than the Heisenberg chain interaction 
as much as possible 
or to separate the intrinsic gap in some skillful way. 
Even in the presence of some difficulties, 
experimental attempts for large-$S$ Haldane gaps 
should be tackled in the future. 

%\begin{acknowledgment}
%\acknowledgment
\section*{Acknowledgments}

We wish to thank Dr. H.~Tadano
for fruitful discussions.
This work was partly supported 
by JSPS KAKENHI Grant Numbers 
16K05418, 16K05419, 16H01080 (JPhysics), and 18H04330 (JPhysics). 
In this research, we used the computational resources of the K computer 
provided by the RIKEN Advanced Institute for Computational Science 
through the HPCI System Research projects 
(Project ID: hp170018, hp170028, hp170070, and hp190053). 
We used the computational resources 
of Fujitsu PRIMERGY CX600M1/CX1640M1 (Oakforest-PACS) 
provided by the 
Joint Center for Advanced High Performance Computing %(JCAHPC) 
through the HPCI System Research project 
(Project ID: hp170207, hp180053, and hp190041). 
Some of the computations were 
performed using the facilities of 
the Department of Simulation Science, 
National Institute for Fusion Science; 
Institute for Solid State Physics, The University of Tokyo;  
and Supercomputing Division, 
Information Technology Center, The University of Tokyo. 

%\end{acknowledgment}

%\appendix
%\section{}
%
%Use the \verb|\appendix| command if you need an appendix(es). The \verb|\section| command should follow even though there is no title for the appendix (see above in the source of this file).
%
%For authors of Invited Review Papers, the \verb|profile| command is prepared for the author(s)' profile.  A simple example is shown below.
%
%\begin{verbatim}
%\profile{Taro Butsuri}{was born in Tokyo, Japan in 1965. ...}
%\end{verbatim}


\begin{thebibliography}{9}
\bibitem{comment_large_scale_job}
The largest scale job was carried out on the K computer, 
using 80,526 nodes, corresponding to 
approximately 97\% of the nodes in the K computer.
\end{thebibliography}
\end{document}